\documentclass[a4paper]{amsart}
\usepackage{epic,eepic,amsmath,amsthm,amssymb,epsfig,cite,indentfirst}
\usepackage{multicol,boxedminipage}

\begin{document}

\title[DW $so(n)$ partition functions]
{On the domain wall partition functions of level-1 affine 
$so(n)$ vertex models}

\author[Dow]{A Dow}
\address{Department of Physics, 
         University of Melbourne,
         Parkville, Victoria 3010, Australia.}
\email{a.dow@ugrad.unimelb.edu.au}

\author[Foda]{O Foda}

\address{Department of Mathematics and Statistics,
         University of Melbourne, 
	 Parkville, Victoria 3010, Australia.}
\email{foda@ms.unimelb.edu.au}

\keywords{Lattice systems, Exactly solvable models, Bethe ansatz} 
\subjclass[2000]{Primary 82B20, 82B23}
\date{}

\newcommand{\field}[1]{\mathbb{#1}}
\newcommand{\C}{\field{C}}
\newcommand{\N}{\field{N}}
\newcommand{\Z}{\field{Z}}

\begin{abstract}
We derive determinant expressions for domain wall partition functions 
of level-1 affine $so(n)$ vertex models, $n\geq 4$, at discrete values 
of the crossing parameter $\lambda$ $=$ $ \frac{m\pi}{2(n-3)}$, 
$m \in \Z$, in the critical regime. 
\end{abstract}

\maketitle

\newtheorem{ca}{Figure}

\def\ll{\left\lgroup}
\def\rr{\right\rgroup}

\newcommand{\Proof}{\medskip\noindent {\it Proof: }}

\def\beqa{\begin{eqnarray}}
\def\eeqa{\end{eqnarray}}
\def\ba{\begin{array}}
\def\ea{\end{array}}
\def\r{\rangle}
\def\l{\langle}
\def\a{\alpha}
\def\b{\beta}
\def\hb{\hat\beta}
\def\d{\delta}
\def\g{\gamma}
\def\e{\epsilon}
\def\tg{\operatorname{tg}}
\def\ctg{\operatorname{ctg}}
\def\sh{\operatorname{sh}}
\def\ch{\operatorname{ch}}
\def\cth{\operatorname{cth}}
\def\th{\operatorname{th}}
\def\eps{\varepsilon}
\def\la{\lambda}
\def\tla{\tilde{\lambda}}
\def\tmu{\tilde{\mu}}
\def\s{\sigma}
\def\sul{\sum\limits}
\def\pl{\prod\limits}
\def\lt({\left(}
\def\rt){\right)}
\def\pd #1{\frac{\partial}{\partial #1}}
\def\const{{\rm const}}
\def\argum{\{\mu_j\},\{\la_k\}} 
\def\umarg{\{\la_k\},\{\mu_j\}} 
\def\prodmu #1{\prod\limits_{j #1 k} \sinh(\mu_k-\mu_j)}
\def\prodla #1{\prod\limits_{j #1 k} \sinh(\lambda_k-\lambda_j)}
\newcommand{\bl}[1]{\makebox[#1em]{}}
\def\tr{\operatorname{tr}}
\def\Res{\operatorname{Res}}
\def\det{\operatorname{det}}

\newcommand{\boldN}{\boldsymbol{N}}
\newcommand{\bra}[1]{\langle\,#1\,|}
\newcommand{\ket}[1]{|\,#1\,\rangle}
\newcommand{\bracket}[1]{\langle\,#1\,\rangle}
\newcommand{\infinity}{\infty}

\renewcommand{\labelenumi}{\S\theenumi.}

\let\up=\uparrow
\let\down=\downarrow
\let\tend=\rightarrow
\hyphenation{boson-ic
             ferm-ion-ic
             para-ferm-ion-ic
             two-dim-ension-al
             two-dim-ension-al
	     rep-resent-ative}

\newtheorem{Theorem}{Theorem}[section]
\newtheorem{Corollary}[Theorem]{Corollary}
\newtheorem{Proposition}[Theorem]{Proposition}
\newtheorem{Conjecture}[Theorem]{Conjecture}
\newtheorem{Lemma}[Theorem]{Lemma}
\newtheorem{Example}[Theorem]{Example}
\newtheorem{Note}[Theorem]{Note}
\newtheorem{Definition}[Theorem]{Definition}
                                                                               
\renewcommand{\mod}{\textup{mod}\,}
\newcommand{\wt}{\text{wt}\,}

\newcommand{\T}{{\mathcal T}}
\newcommand{\U}{{\mathcal U}}
\newcommand{\tT}{\tilde{\mathcal T}}
\newcommand{\tU}{\widetilde{\mathcal U}}
\newcommand{\Y}{{\mathcal Y}}
\newcommand{\B}{{\mathcal B}}
\newcommand{\D}{{\mathcal D}}
\newcommand{\M}{{\mathcal M}}
\renewcommand{\P}{{\mathcal P}}
\newcommand{\R}{{\mathcal R}}

\hyphenation{And-rews
             Gor-don
             boson-ic
             ferm-ion-ic
             para-ferm-ion-ic
             two-dim-ension-al
             two-dim-ension-al}

\setcounter{section}{-1}

\section{Introduction}\label{introduction}

In \cite{korepin}, Korepin introduced domain wall (DW) boundary 
conditions in the context of the six vertex, spin-$1/2$ or level-1 
$A^{(1)}_1$ model on a finite lattice, and obtained recursion 
relations that determine the partition function in that case. In 
\cite{izergin}, Izergin solved Korepin's recursion relations and 
obtained a determinant expression for the level-1 $A^{(1)}_1$ DW 
partition function.

In \cite{cfk}, determinant expressions were obtained for the spin-$k/2$, 
or level-$k$ $A^{(1)}_1$ DW partition functions, $k \in \N$, using the 
fact that these models can be obtained from the level-1 model using 
fusion \cite{rks, djkmo}. Fusion was necessary in proving the level-$k$ 
$A^{(1)}_1$ result because an Izergin type proof, based on Lagrange 
interpolation, fails to extend for general $k$. 

The reason is that the DW spin-$k/2$ partition function is a polynomial 
of degree $k(L-1)$ in each (multiplicative) rapidity variable (where 
\lq degree\rq is defined as the difference of the highest and lowest 
powers in the variable). Thus for Lagrange interpolation to work, one 
needs to impose $k(L-1)+1$ conditions on the partition function, which 
are not available for $k>2$ because the DW boundary conditions offer 
exactly $2L$ such conditions ($L$ from the upper left corner and $L$ 
from the upper right corner, while the lower corners offer no new 
conditions due to the symmetries of the lattice and the vertex weights).

For $k=2$, there is an `extended' Izergin type proof, in addition to the 
fusion proof. This proof was used, in \cite{cfk}, as a check, in the 
special $k=2$ case, on the general $k$ fusion proof. However, strictly
speaking, this proof was unnecessary and the extra $L$ conditions
could have remained unused. This is intriguing, since exactly solvable 
models are notoriously parsimonious. In this work we find an application 
for them.

Since the 19 vertex, spin-1, or level-2 $A^{(1)}_1$ model coincides 
with the level-1 affine $so(3)$ vertex model (as can be seen by 
comparing the vertex weights), and because the degree of the vertex 
weights of level-1 affine $so(n)$ models, as polynomials in the 
rapidities, is independent of $n$ \cite{jimbo}, it is natural 
to expect that the DW partition functions of these models have the 
same form as that of the level-2 $A^{(1)}_1$ model.

In this work, we propose a determinant expression for the DW partition 
functions of level-1 affine $so(n)$ vertex models\footnote{For simplicity, 
we refer to these models from now on as the $so(n)$ vertex models.}, 
for $n\geq 4$, based on the result of \cite{cfk} for the level-2 
$A^{(1)}_1$ case. 

Since the $so(n)$ models, for $n\geq 4$, cannot be obtained by fusing 
simpler models, we need to use $2L-1$ of the $2L$ conditions provided 
by DW boundaries to obtain an Izergin type proof. We find that our 
proposal is correct, but only at discrete values of the crossing parameter 
$\lambda$, namely $\lambda = \frac{m\pi}{2(n-3)}$, $m \in \Z$, in
the critical regime (trigonometric vertex weightsi for real
rapidity variables). For $n=3$, 
one has the level-2 $A^{(1)}_1$ result, valid for all $\lambda$. We 
have no insight into this restriction on $\lambda$, and the situation 
for $n\geq 4$ and general $\lambda$ is open.

In section \ref{vertex-models}, we briefly recall the $so(n)$ vertex 
models, define what we mean by DW boundary conditions in this context, 
and propose a determinant expression for the DW partition functions, 
that initially depends on continuous $\lambda$.

In section \ref{proof}, we outline an Izergin type proof of the proposed 
determinant expression, show that it is valid only at 
$\lambda = \frac{m \pi}{2(n-3)}$, $m \in \Z$, and comment on the
the very little that we know about DW partition functions of vertex models 
based on other semi-simple Lie algebras. The presentation is elementary, 
and (almost) self contained.

\section{Determinant partition functions}\label{vertex-models}

\subsection{Definitions}

We work on a square lattice consisting of $L$ horizontal lines 
(labelled from bottom to top), and $L$ vertical lines (labelled 
from left to right). 

We assign the {\it i}-th horizontal line an orientation from left 
to right, and a complex rapidity variable $x_i$. We assign the 
{\it j}-th vertical line an orientation from bottom to top, and 
a complex rapidity variable $y_j$. 

Each horizontal (vertical) line intersects with $L$ vertical 
(horizontal) lines. There are $L^2$ intersection points. A line
segment attached to two intersection points is a bond. A segment 
attached to one intersection point is a boundary bond.

%
\begin{center}
\begin{minipage}{4.9in}
\setlength{\unitlength}{0.0008cm}
\begin{picture}(4800, 6000)(-100, 0)
\thicklines
\path(2400,5400)(2400,1800)
\path(3000,5400)(3000,1800)
\path(3600,5400)(3600,1800)
\path(4200,5400)(4200,1800)
\path(4800,5400)(4800,1800)
\path(1800,4800)(5400,4800)
\path(1800,4200)(5400,4200)
\path(1800,3600)(5400,3600)
\path(1800,3000)(5400,3000)
\path(1800,2400)(5400,2400)
\path(0600,4254)(1200,4254)
\path(2400,654)(2400,1254)
\path(3000,654)(3000,1254)
\path(3600,654)(3600,1254)
\path(4200,654)(4200,1254)
\path(4800,654)(4800,1254)
\path(600,2454)(1200,2454)
\path(600,3054)(1200,3054)
\path(600,3654)(1200,3654)
\path(600,4854)(1200,4854)
\whiten\path(2490,894)(2400,1254)(2310,894)(2490,894)
\whiten\path(3090,894)(3000,1254)(2910,894)(3090,894)
\whiten\path(3690,894)(3600,1254)(3510,894)(3690,894)
\whiten\path(4290,894)(4200,1254)(4110,894)(4290,894)
\whiten\path(4890,894)(4800,1254)(4710,894)(4890,894)
\whiten\path(840,2364)(1200,2454)(840,2544)(840,2364)
\whiten\path(840,2964)(1200,3054)(840,3144)(840,2964)
\whiten\path(840,3564)(1200,3654)(840,3744)(840,3564)
\whiten\path(840,4164)(1200,4254)(840,4344)(840,4164)
\whiten\path(840,4764)(1200,4854)(840,4944)(840,4764)
\put(-100,4854){$x_L$}
\put(-100,3054){$x_2$}
\put(-100,2454){$x_1$}
\put(2300,0250){$y_1$}
\put(2900,0250){$y_2$}
\put(4700,0250){$y_L$}
\end{picture}
\begin{ca} 
\label{lattice}
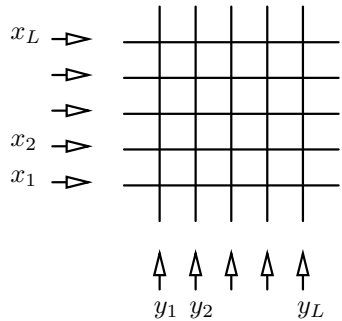
A finite square lattice, with oriented lines and rapidities. 
\end{ca}
\end{minipage}
\end{center}

\bigskip

The intersection of the $i$-th horizontal line and the $j$-th 
vertical line, together with the 4 bonds attached to it, and 
the variables on these bonds, is a vertex $v_{ij}$. 

To each vertex $v_{ij}$, we assign a weight $w_{ij}$, that 
depends on the difference of rapidity variables flowing through 
that vertex. In exactly solvable models, the weights satisfy the 
Yang Baxter equations \cite{baxter-book}.

The partition function of a vertex model, on an $L\times L$ lattice, 
$Z^{}_{L\times L}$, is a weighted sum over all configurations, that 
satisfy certain boundary conditions. The weight of a configuration 
is the product of the weights, $w_{ij}$, of the vertices $v_{ij}$. 

\begin{equation}
Z^{}_{L\times L}\ll \{{\bf x}\}, \{{\bf y}\}\rr = 
\sul_{\begin{subarray}{c}
      {\text config-}\\
      {\text urations }
        \end{subarray}}
\ll \pl_{\rm vertices} w_{ij} \rr
\label{physical}
\end{equation}

\subsection{Notation}

We use the notation $[x]= \sin(\lambda x)$, where $\lambda$ is the 
crossing parameter. We also use ${\bf \langle x\rangle}= 1/[x]$. 

$E_{ij}$ is the $n\times n$ matrix with all elements equal to $0$, 
except the $ij$-th element, which is equal to 1. $n$ is the dimension 
of the vector representation of $so(n)$, 
the dynamical variables are colours labeled by an index 
$\alpha \in \{1, 2, \cdots, n\}$, each colour $\alpha$ has a
conjugate colour labelled by $\alpha^{\prime}$, where
$\alpha^{\prime}  + \alpha = n+1$, and 

\begin{equation}
\bar{\alpha} =
\begin{cases}
\alpha + \frac{1}{2}, &           \alpha  <    \frac{n+1}{2}\\
\alpha,               &           \alpha  =    \frac{n+1}{2}\\
\alpha - \frac{1}{2}, &           \alpha  >    \frac{n+1}{2} 
\end{cases}
\end{equation}

Finally, we define $\tilde{\alpha} = \alpha - \alpha^{\prime}$,
so that

\begin{equation}
\tilde{\alpha} =
\begin{cases}
- \frac{1}{2}, &           \alpha  <    \frac{n+1}{2}\\
0,             &           \alpha  =    \frac{n+1}{2}\\
+ \frac{1}{2}, &           \alpha  >    \frac{n+1}{2}
\end{cases}
\end{equation}

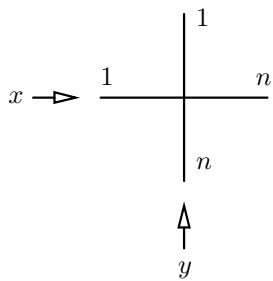
\subsection{The $so(n)$ vertex models}

In the $so(n)$ vertex model, each bond is assigned a colour 
$\kappa$ $\in$ $\{1, 2, \cdots, n\}$. The allowed interactions,
and their dependences on the rapidity variables are coded in the
quantum $R$ matrix. From \cite{jimbo}, the quantum $R$ matrix of 
level-1 affine $so(n)$ vertex models is

\begin{alignat}{8}
\label{r}
R(u) 
&=& 
&\ \ w_1& 
&\sul_{\alpha \neq {\alpha^{\prime}}}& 
&E_{\alpha \alpha} \otimes E_{\alpha \alpha}& 
&+&
&\ \ w_2&
&\sul_{\begin{subarray}{c}
                            \alpha \neq  \beta \\
                            \alpha \neq {\beta^{\prime}}
\end{subarray}}&
&E_{\alpha \alpha} \otimes E_{\beta \beta} \\
&\ \ +& 
&\ \ w_3&
&\sul_{\alpha \neq \alpha^\prime}&
&E_{\alpha \alpha} \otimes E_{{\alpha^\prime} {\alpha^\prime}}& 
&\ \ +& 
&\ \ w_4&
&\sul_{\alpha = \alpha^\prime}&
&E_{\alpha \alpha} \otimes E_{{\alpha} {\alpha}}\\ \nonumber
&\ \ +&
&\ \ &
&\sul_{\begin{subarray}{c}
                           \alpha <     \beta \\
                           \alpha \neq {\beta^\prime}
\end{subarray}}&
& w_5^{(\alpha \beta)} E_{\alpha \beta} \otimes E_{\beta \alpha}& 
&+&
& \ \ &
&\sul_{\begin{subarray}{c}
                           \alpha >     \beta \\
                           \alpha \neq {\beta^\prime}
\end{subarray}}&
& w_6^{(\alpha \beta)} E_{\alpha \beta} \otimes E_{\beta \alpha} \\ \nonumber
&\ \ +&
&\ \ &
&\sul_{\alpha < \beta}&
& w_7^{(\alpha \beta)} 
E_{\alpha \beta} \otimes E_{{\alpha^\prime} {\beta^\prime}} &
&\ \ +&
& \ \ &
&\sul_{\alpha > \beta}&
& w_8^{(\alpha \beta)} 
E_{\alpha \beta} \otimes E_{{\alpha^\prime} {\beta^\prime}} 
\nonumber
\label{r}
\end{alignat}

The vertex (Boltzmann) weights are typically written either in 
multiplicative notation, as in \cite{jimbo}, so that the weights 
depend on ratio, $U = \frac{Y}{X}$ of the {\it multiplicative 
rapidity variables}, $X$ and $Y$, or in additive notation, so 
that the weights depend on the difference $u = -x+y$ of the {\it 
additive rapidity variables}, $x$ and $y$. These variables are 
related by $X = k^{x}$, {\it etc}, where $k=\exp(-i 2\lambda)$, 
and $\lambda$ is the crossing parameter. We use either notation 
depending on which is more convenient.

\subsubsection{Multiplicative notation} 

\begin{align*}
w_1(U) &=  (U-k^{n-2})(U-k^2),\\ 
w_2(U) &=  (U-k^{n-2})(U-1)k,\\ 
w_3(U) &=  (U-k^{n-4})(U-1)k^2,\\ 
w_4(U) &=  (U-k^{n-2})(U-1)k 
        +  (1 - k^{n-2})(1 - k^2)U, \\ 
w_{5}^{(\alpha \beta)}(U) &=   (U -k^{n-2})(1-k^2)
                                U^{-\tilde{\alpha} +
				\tilde{\beta}},\\
w_{6}^{(\alpha \beta)}(U) &=   (U -k^{n-2})(1-k^2)
                                U^{-\tilde{\alpha}+\tilde{\beta}+1},\\ 
w_{7}^{(\alpha \beta)}(U) &=   
\left\lgroup
(U-k^{n-2}) \delta_{\alpha \beta^\prime} 
	- (U-1) k^{\bar{\alpha} - \bar{\beta} + n-2} 
\right\rgroup (1 - k^2) 
U^{-\tilde{\alpha} + \tilde{\beta}},\\
w_{8}^{(\alpha \beta)}(U) &=   
\left\lgroup
(U-k^{n-2}) \delta_{\alpha \beta^\prime}   
	- (U-1) k^{\bar{\alpha} - \bar{\beta} + n-2} 
\right\rgroup (1-k^2) 
U^{- \tilde{\alpha} + \tilde{\beta} + 1}
\end{align*}

\subsubsection{Additive notation} 

\begin{align*}
w_1(u) &=  [u + n-2][u + 2],\\
w_2(u) &=  [u + n-2][u],\\
w_3(u) &=  [u + n-4][u],\\
w_4(u) &=  [u + n-2][u] + [2][n-2],\\
w_{5}^{(\alpha \beta)}(u) 
&=  [u + n-2][2] e^{- i \lambda u} \phi(u, \alpha, \beta),\\
w_{6}^{(\alpha \beta)}(u) 
&=  [u + n-2][2] e^{+ i \lambda u} \phi(u, \alpha, \beta),\\
w_{7}^{(\alpha \beta)}(u) 
&=  
\left\lgroup 
    [u + n-2] \delta_{\alpha \beta^\prime} 
 -  [u] e^{i \lambda (2\bar{\beta} - 2 \bar{\alpha} - n + 2)}
    \right\rgroup [2] e^{- i \lambda u} \phi(u, \alpha, \beta), \\
w_{8}^{(\alpha \beta)}(u) 
&=  
\left\lgroup 
    [u+n-2] \delta_{\alpha \beta^\prime} 
 -  [u]  e^{i \lambda (2\bar{\beta} - 2\bar{\alpha} - n + 2)} 
    \right\rgroup [2] e^{+i \lambda u} \phi(u, \alpha, \beta) 
\end{align*}

\noindent where $\phi{(u, \alpha, \beta)} = 
\exp 
\left\lgroup 
2 i \lambda (\tilde{\alpha} - \tilde{\beta}) u 
\right\rgroup
$ 
Further, we have removed an overall factor of 
$e^{i \lambda(2u - n)}$ from the additive vertex weights.

\subsection{Symmetrizing factors} 

The purpose of the factors of $U^{-\tilde{\alpha} + \tilde{\beta}}$
in the multiplicative notation, or $\phi{(u, \alpha, \beta)}$ in
the additive notation (and which are not present in \cite{jimbo})
is to Symmetrize the vertices in a way that simplifies our calculations. 
The Yang Baxter equations remain unchanged by such factors. They appear 
only in the weights $\{w_5, w_6, w_7, w_8\}$, as  they reduce to 1 for 
$\{w_1, w_2, w_3, w_4\}$ are 1. 

\subsection{Degrees} We define the degree of a vertex weight,
as a polynomial in a (multiplicative) rapidity variable as the
difference of the highest power and lowest power of terms
appearing in the polynomial. Thus the degree of each of $\{w_1, w_2,
w_3, w_4\}$ is 2, the degree of each of $\{w_5, w_6\}$ is 1, and
the degree of each of $\{w_7, w_8\}$ is either 1 or 0 depending 
on the colours flowing. 

The degree of the DW partition function as a polynomial in a
(multiplicative) rapidity variable is the sum of the degrees of
the vertex weights that depend on that variable. In the additive
notation, one can also compute the same degree, but now for
trigonometric polynomials.

\subsection{Scattering of colours} One can think of an $so(n)$ 
vertex as an elastic scattering of two particles that come in 
from the left and from below (same directions as the incoming 
rapidities), then exit to the right and to the top (same directions 
as the outgoing rapidities). 

A term proportional to $E_{\rho \sigma} \otimes E_{\mu \nu}$, on 
the RHS of equation \ref{r}, describes the elastic scattering of 
a horizontally moving particle, with rapidity $x$, that comes in 
from the left with colour $\rho$ and exists to the right with 
colour $\sigma$, and a vertically moving particle, with rapidity 
$y$, that comes in from below with colour $\mu$ and exists to the 
top, with colour $\nu$. The strength of the interaction, coded in 
the weight $w$, depends on the rapidity difference $-x+y$, and on 
the crossing parameter $\lambda$, and is non vanishing only if 
$\rho + \mu = \sigma + \nu$.

Notice that an inflowing colour $\alpha$, must either exit the 
interaction (in the same direction, or after changing directions), 
or annihilate with an incoming conjugate colour $\alpha^{\prime}$, 
and a pair of conjugate colours, $\beta$ and $\beta^{\prime}$ exit 
the interaction.

%
\begin{center}
\begin{minipage}{4.9in}
\setlength{\unitlength}{0.0008cm}
%


\begin{picture}(4000, 5000)(-0920,-1000)

\thicklines


\path(0620,2220)(3420,2220)
 \put(-920,2120){$x$}
\whiten\path(-160,2130)(0200,2220)(-160,2310)(-160,2130)
\path(-520,2220)(-160,2220)
\put(0620,2420){$\rho$}
\put(3220,2420){$\sigma$}


\path(2020,0820)(2020,3620)
\whiten\path(2110,40)(2020,400)(1930,40)(2110,40)
\path(2020,-320)(2020,40)
\put(1920,-720){$y$}
\put(2220,1020){$\mu$}
\put(2220,3420){$\nu$}
\end{picture}
\begin{ca}
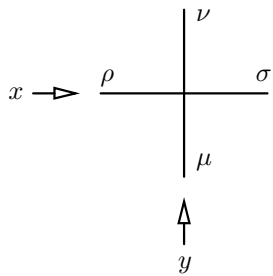
A general $so(n)$ vertex, of weight $w(-x+y)$ corresponding 
to a term $w(-x+y) E_{\rho \sigma}\otimes E_{\mu \nu}$ in the 
quantum $R(-x+y)$ matrix. 
\label{general-vertex}
\end{ca}
\end{minipage}
\end{center}

\bigskip

\subsection{Minimal and maximal colours} 
We refer to the colour with index $1$ as minimal. 
We refer to the colour with index $n$ as maximal. 

\subsection{The $c+$ vertex} We refer to the unique vertex 
with minimal colour incoming from the left, 
maximal colour exiting to the right, 
maximal colour incoming from below, and 
minimal colour exiting to the top, as shown in
Figure \ref{c+vertex}, as the $c+$ vertex. 

%
\begin{center}
\begin{minipage}{4.9in}
\setlength{\unitlength}{0.0008cm}
\begin{picture}(4000, 5000)(-0920,-1000)
\thicklines


\path(0620,2220)(3420,2220)
\put(-920,2120){$x$}
\whiten\path(-160,2130)(0200,2220)(-160,2310)(-160,2130)
\path(-520,2220)(-160,2220)
\put(0620,2420){$1$}
\put(3220,2420){$n$}


\path(2020,0820)(2020,3620)
\whiten\path(2110,40)(2020,400)(1930,40)(2110,40)
\path(2020,-320)(2020,40)
\put(1920,-720){$y$}
\put(2220,1020){$n$}
\put(2220,3420){$1$}
\end{picture}
\begin{ca}
The $so(n)$ $c+$ vertex.
\label{c+vertex}
\end{ca}
\end{minipage}
\end{center}

\bigskip

\subsection{Domain wall (DW) boundary conditions} Consider the $so(n)$ 
vertex model on a finite square lattice, and require that the boundary 
bonds have the same colours as the bond on the corresponding boundary 
of the $c+$ vertex: 
left  and upper boundary bonds carry minimal colours, 
right and lower boundary bonds carry maximal colours. 

%
\begin{center}
\begin{minipage}{4.9in}
\setlength{\unitlength}{0.0008cm}
\begin{picture}(4800, 6000)(0,-100)
\thicklines


\path(2400,5400)(2400,1800)
 \put(2100,5400){$1$}
 \put(2100,1600){$n$}

\path(3000,5400)(3000,1800)
 \put(2700,5400){$1$}
 \put(2700,1600){$n$}

\path(3600,5400)(3600,1800)
 \put(3300,5400){$1$}
 \put(3300,1600){$n$}

\path(4200,5400)(4200,1800)
 \put(3900,5400){$1$}
 \put(3900,1600){$n$}

\path(4800,5400)(4800,1800)
 \put(4500,5400){$1$}
 \put(4500,1600){$n$}

\path(1800,4800)(5400,4800)
 \put(1800,4900){$1$}
 \put(5400,4900){$n$}

\path(1800,4200)(5400,4200)
 \put(1800,4300){$1$}
 \put(5400,4300){$n$}

\path(1800,3600)(5400,3600)
 \put(1800,3700){$1$}
 \put(5400,3700){$n$}

\path(1800,3000)(5400,3000)
 \put(1800,3100){$1$}
 \put(5400,3100){$n$}

\path(1800,2400)(5400,2400)
 \put(1800,2500){$1$}
 \put(5400,2500){$n$}

\path(0600,4254)(1200,4254)
\path(2400,0654)(2400,1254)
\path(3000,0654)(3000,1254)
\path(3600,0654)(3600,1254)
\path(4200,0654)(4200,1254)
\path(4800,0654)(4800,1254)
\path(0600,2454)(1200,2454)
\path(0600,3054)(1200,3054)
\path(0600,3654)(1200,3654)
\path(0600,4854)(1200,4854)
\whiten\path(2490,894)(2400,1254)(2310,894)(2490,894)
\whiten\path(3090,894)(3000,1254)(2910,894)(3090,894)
\whiten\path(3690,894)(3600,1254)(3510,894)(3690,894)
\whiten\path(4290,894)(4200,1254)(4110,894)(4290,894)
\whiten\path(4890,894)(4800,1254)(4710,894)(4890,894)
\whiten\path(840,2364)(1200,2454)(840,2544)(840,2364)
\whiten\path(840,2964)(1200,3054)(840,3144)(840,2964)
\whiten\path(840,3564)(1200,3654)(840,3744)(840,3564)
\whiten\path(840,4164)(1200,4254)(840,4344)(840,4164)
\whiten\path(840,4764)(1200,4854)(840,4944)(840,4764)
\put(-100,4854){$x_L$}
\put(-100,3054){$x_2$}
\put(-100,2454){$x_1$}
\put(2300,0250){$y_1$}
\put(2900,0250){$y_2$}
\put(4700,0250){$y_L$}
\end{picture}

\begin{ca}
\label{dwbc}
$so(n)$ DW boundary conditions.
\end{ca}

\end{minipage}
\end{center}

\bigskip

\subsection{Determinant DW partition functions}

Based on the result for spin-1 DW partition function in
\cite{cfk}, we propose the following expression for the 
$so(n)$ DW partition function

\bigskip 
\begin{boxedminipage}[c]{12cm}

\begin{equation}
Z^{(n)}_{L\times L} \ll \{{\bf x}\},\{{\bf y}\} \rr
=  
\frac{N}{D} \times det \ll M^{2\times 2}_{2L\times 2L}\rr, 
\label{proposal}
\end{equation}

\noindent where
$$
N = 
\pl_{1 \le i, j \le L} w_1(-x_i+ y_j) w_2(-x_i+ y_j) w_3(-x_i+ y_j), 
$$
$$
D = 
\pl_{i < j} w_2(-x_i+ x_j) w_3(-x_i+ x_j) w_2(y_j- y_i) w_3(y_j- y_i),
$$

\begin{align}
\begin{array}{lll}
M_{2j-1,             2i-1}^{         }             ={1}/{w_2(-x_i+y_j)},&
M_{2j-1,             2i{\phantom{-1}}}^{         } ={1}/{w_1(-x_i+y_j)},\\
\phantom{space}&\phantom{space}\\
M_{2j,{\phantom{-1}} 2i-1}^{         }             ={1}/{w_3(-x_i+y_j)},&
M_{2j,{\phantom{-1}} 2i{\phantom{-1}}}^{         } ={1}/{w_2(-x_i+y_j)}
\nonumber
\end{array}
\end{align}

\end{boxedminipage}
\bigskip

At this stage, we have no restriction on the crossing parameter
$\lambda$. We will see shortly that equation \ref{proposal} is 
valid only for 

\bigskip
\begin{boxedminipage}[c]{6cm}
\begin{equation}
\lambda = \frac{m \pi}{2(n-3)}, \quad m \in \Z
\label{condition}
\end{equation}
\end{boxedminipage}

\bigskip
\section{Proof}\label{proof}

We prove equation \ref{proposal} following the same arguments as in
\cite{izergin}, suitably extended as in \cite{cfk}. Using the
abbreviations

$$
{\rm LHS} =  Z^{(n)}_{L\times L} \ll \{{\bf x}\},\{{\bf y}\} \rr,
\quad
{\rm RHS} =  \frac{N}{D} \times det \ll M^{2\times 2}_{2L\times 2L}\rr,
$$
conceptually, one needs to show that each side is 

\begin{enumerate}

\item A symmetric function in the $\{ {\bf x} \}$ variables, and in 
the $\{ {\bf y} \}$ variables. This allows one to choose to work in 
terms of any single $x$, or single $y$ rapidity variable, with no loss 
of generality. 

\item A polynomial in the rapidity variables (rather than a rational 
function). This allows one to use Lagrange interpolation to determine 
it. 

\item A polynomial in any one rapidity variable (keeping all others 
fixed) of degree $2L-2$, where by \lq degree\rq we mean the difference
between the highest and lowest powers of the variable in the
polynomial. 

\item Equal to the other polynomial side at $L$ distinct values 
of the same rapidity variable, as obtained by imposing conditions 
on the rapidities intersecting at the upper left corner.

\item Equal to the other polynomial side at $L$ distinct values 
of the same rapidity variable, as obtained by imposing conditions 
on the rapidities intersecting at the upper right corner.

\item Subject to the same initial condition.

\end{enumerate}

This would imply that the LHS $=$ RHS. Notice that we use $2L$ 
conditions, rather than the $2L-1$ that we need, because we can 
and because that simplifies the discussion.

\subsection{Symmetric functions}

\subsubsection{LHS} It is straightforward to use the Yang Baxter 
equations, just as in Izergin's proof in the six vertex model 
case, to permute any two adjacent lines. For example, to permute
the $i$-th and the $(i+1)$-st horizontal lines, we attach a $w_1$ 
vertex to these two lines from the left. 


\begin{center}
\begin{minipage}{4.9in}
\setlength{\unitlength}{0.0008cm}

\begin{picture}(3000,3000)(-2000, 0)
\thicklines

\path(1222,2722)(1222,0022)
\path(2122,2722)(2122,0022)
\path(2122,1822)(2722,1822)
\path(2122,0922)(2722,0922)
\thinlines
\path(1522,2422)(1222,2122)
\path(1822,2422)(1222,1822)
\path(2122,2422)(1222,1522)
\path(2122,2122)(1222,1222)
\path(2122,1822)(1222,0922)
\path(2122,1522)(1222,0622)
\path(2122,1222)(1222,0322)
\path(2122,0922)(1522,0322)
\path(2122,0622)(1822,0322)

\thicklines
\path(1222,1822)(2122,1822)
\path(1222,0922)(2122,0922)
\path(1222,1822)(0922,1822)
\path(1222,0922)(0922,0922)
\path(0922,0922)(0472,1372)
\path(0922,1822)(0472,1372)

\path(0472,1372)(0022,0922)
\path(0472,1372)(0022,1822)

\put(-2000,1800){$x_{i  }$}

\whiten\path(-0660,1710)(-300,1800)(-0660,1890)(-0660,1710)
\path(-1020,1800)(-0660,1800)

\put(-110,2000){$1$}
\put(0890,2000){$1$}
\put(2610,2000){$n$}
\put(3000,1800){$x_{i+1}$}

\put(-2000,0900){$x_{i+1}$}

\whiten\path(-0660,0810)(-300,0900)(-0660,0990)(-0660,0810)
\path(-1020,0900)(-0660,0900)

\put(-110,0500){$1$}
\put(0890,0500){$1$}
\put(2610,0500){$n$}
\put(3000,0900){$x_{i  }$}

\end{picture}

\begin{ca}
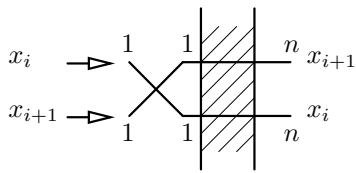
Attaching a vertex with weight $w_1$ from the left.
\end{ca}

\end{minipage}
\end{center}

\bigskip

The augmented partition function now is 

$$
Z_{L\times L}^{\text aug} \ll \{ \cdots, x_i, x_{i+1}, \cdots \} \rr
= 
w_1(-x_{i+1} + x_{i}) 
      {Z}_{L\times L} \ll \{ \cdots, x_i, x_{i+1}, \cdots \} \rr
$$

Using the Yang Baxter equations, we can thread the extra vertex 
all the way to the other side, from which it emerges once again 
as a $w_1$ with the same vertex weight


\begin{center}
\begin{minipage}{4.9in}
\setlength{\unitlength}{0.0008cm}

\begin{picture}(5000,3000)(-1360, 0)
\thicklines
\path(1350,2722)(1350,0022)
\path(2250,2722)(2250,0022)
\path(0750,1822)(2550,1822)(3450,0922)
\path(0750,0922)(2550,0922)(3450,1822)
\path(3075,1297)(3450,0922)
\path(1350,2722)(1350,2272)
\path(1350,0022)(1350,0472)
\path(2250,0022)(2250,0472)
\path(2250,2722)(2250,2272)
\thinlines
\path(1650,2422)(1350,2122)
\path(1650,2422)(1350,2122)
\path(1950,2422)(1350,1822)
\path(1950,2422)(1350,1822)
\path(2250,2422)(1350,1522)
\path(2250,2422)(1350,1522)
\path(2250,2122)(1350,1222)
\path(2250,2122)(1350,1222)
\path(2250,1822)(1350,0922)
\path(2250,1822)(1350,0922)
\path(2250,1522)(1350,0622)
\path(2250,1522)(1350,0622)
\path(2250,1222)(1350,0322)
\path(2250,1222)(1350,0322)
\path(2250,0922)(1650,0322)
\path(2250,0922)(1650,0322)
\path(2250,0622)(1950,0322)
\path(2250,0622)(1950,0322)
\thicklines
\path(0750,1822)(1350,1822)
\path(0750,0922)(1350,0922)

\put(0640,2000){$1$}
\put(2360,2000){$n$}
\put(3360,2000){$n$}

\put(0640,0500){$1$}
\put(2360,0500){$n$}
\put(3360,0500){$n$}

\put(3750,1822){$x_{i+1}$}
\put(3750,0922){$x_{i  }$}

\put(-1360,1800){$x_{i  }$}
\whiten\path(-0020,1710)(0340,1800)(-0020,1890)(-0020,1710)
       \path(-0380,1800)(-020,1800)

\put(-1360,0900){$x_{i+1}$}
\whiten\path(-0020,0810)(0340,0900)(-0020,0990)(-0020,0810)
       \path(-0380,0900)(-020,0900)

\end{picture}

\begin{ca}
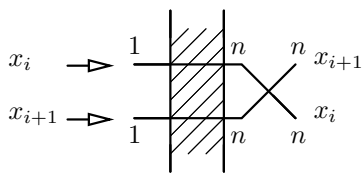
Extracting a vertex with weight $w_1$ from the right.
  \end{ca}

\end{minipage}
\end{center}

\bigskip
\noindent and the augmented partition function becomes 

$$
Z_{L\times L}^{\text aug} \ll \{ \cdots, x_i, x_{i+1}, \cdots \} \rr
=
      {Z}_{L\times L} \ll \{ \cdots, x_{i+1}, x_i, \cdots \} \rr
w_1(-x_{i+1} + x_{i})
$$

\noindent from which we conclude that 
$$
{Z}_{L\times L} \ll \{ \cdots, x_i, x_{i+1}, \cdots \} \rr =
{Z}_{L\times L} \ll \{ \cdots, x_{i+1}, x_i, \cdots \} \rr
$$

\subsubsection{RHS} We can check the symmetry of the determinant 
expression by direct calculation. $N$ is symmetric by construction 
under exchange of any pair of $x$ rapidity variables, or $y$ rapidity 
variables. The determinant is also symmetric under the same changes 
due to the $2\times 2$ block structure. 

It is straightforward to show that the denominator $D$ is also 
symmetric, but only when one uses the periodicity that follow 
from choosing $\lambda$ $=$ $\frac{m \pi}{2(n-3)}$, $m \in \Z$.
This goes as follows: 

We wish to check whether
$D = \pl_{i < j} w_2(-x_i+ x_j) w_3(-x_i+ x_j) 
                 w_2(y_j- y_i) w_3(y_j- y_i)$ 
is symmetric under exchanging $\{x_{\rho}, x_{\sigma}\}$, where 
$1 \leq \rho < \sigma \leq n$. It is sufficient to consider the 
$x$-dependent factors, 

$$
\ll \pl_{i < j} [-x_i + x_j]^2 \rr
\ll \pl_{i < j} [-x_i + x_j + n - 2] [-x_i + x_j + n - 4] \rr
$$ 

The first product is manifestly invariant. In the second product, 
there are $2(\sigma - \rho) - 1$ (double) factors that will change,
and can be re-written as  
$[-x_i + x_j - n+2] [-x_i + x_j -n+4]$, 
where in $(\sigma - \rho)$ cases, 
$\{i = \rho, j \neq \sigma\}$, 
in $(\sigma - \rho)$ cases, 
$\{i \neq \rho, j = \sigma\}$, 
and in one case $\{i = \rho, j = \sigma\}$.
The only way to restore invariance is to restrict the crossing
parameter $\lambda$ according to equation \ref{condition}.

This is the first indication that we need the above constraint 
on $\lambda$. The same condition turns out to be necessary at 
almost every subsequent step of the proof.

\subsection{Symmetric polynomials}

\subsubsection{LHS} The physical partition function is, by
construction, a polynomial in each (multiplicative) rapidity 
variable and has no poles at finite points in the complex 
plane.  

\subsubsection{RHS} The determinant form has poles that shouldn't 
be there, because the denominator product 
$
D = \pl_{i < j} 
w_2(-x_i+ x_j) w_3(-x_i+ x_j) 
w_2( y_j- y_i) w_3( y_j- y_i)
$
has double zeros for $-x_i + x_j = 0$, 
simple zeros for $-x_i + x_j + n-2=0$ and for $-x_i + x_j + n-4=0$, 
and similarly for the $y$ variables. These poles must cancel. 
They cannot be cancelled by zeros in the numerator product 
$N = 
\pl_{i \le i, j \le L} w_1(-x_i+ y_j) w_2(-x_i+ y_j) w_3(-x_i+ y_j)$. They 
must be cancelled by zeros of the determinant. Due to the $2\times 2$ 
structure of the determinant, it is clear that the double poles at 
$-x_i + x_j = 0$ are cancelled by double zeros of the determinant (the 
double $i$-th rows become equal to the double $j$-th rows).

The cancellations of the simple poles at $-x_i + x_j + n-2=0$ and 
$-x_i + x_j + n-4=0$ (and at corresponding values of the {\bf $y$} 
variables) is not automatic. Consider the $i$-th and $j$-th double rows 
of a generic $k$-th column.

\begin{align}
\ll
\begin{array}{lllll}
&\vdots&\vdots&\vdots& \\
\ldots&
\langle -x_i+y_k      \rangle \langle -x_i+y_k+n-2    \rangle&
\langle -x_i+y_k+ n-2 \rangle \langle -x_i+y_k+2       \rangle&
\ldots& \\
\ldots&
\langle -x_i+y_k      \rangle \langle -x_i+y_k+n-4\rangle&
\langle -x_i+y_k      \rangle \langle -x_i+y_k+n-2    \rangle&
\ldots& \\ 
&\vdots&\vdots&\vdots&                                   \\
\ldots&
\langle -x_j+y_k      \rangle \langle -x_j+y_k+n-2    \rangle& 
\langle -x_j+y_k+n-2 \rangle \langle -x_j+y_k+2       \rangle&
\ldots& \\
\ldots&
\langle -x_j+y_k      \rangle \langle -x_j+y_k+n-4\rangle&
\langle -x_j+y_k      \rangle \langle -x_j+y_k+n-2\rangle&
\ldots& \\
&\vdots&\vdots&\vdots& 
\end{array}
\rr
\nonumber
\end{align}

We find that we can achieve cancellation only if equation
\ref{condition} is satisfied.

When this is the case, the vertex weights are such that, 
for $-x_i + x_j + n-2=0$,
the upper $i$-th row is equal (up to a minus sign for odd
values of $m$ in condition \ref{condition}) to 
the lower $j$-th row for $-x_i + x_j + n-2=0$, 
and 
for $-x_i + x_j + n-4=0$,
the lower $i$-th row 
is equal (up to a minus sign for odd values of $m$ in condition
\ref{condition}) to 
the upper $j$-th row. 

For $n=3$ (the 19 vertex model), the above relations are automatically 
satisfied, and one does not need to impose an extra condition on 
$\lambda$.  

\subsection{Symmetric polynomials of degree $2L-2$} 

\subsubsection{LHS} To determine the degree of the partition function 
as a polynomial in any horizontal or vertical rapidity variable, it 
is sufficient to consider any one variable from a certain type, and 
use symmetry to conclude that the result applies to all other variables 
of the same type. In the following we consider $x_L$, and refer to it, 
for simplicity, as $x$. It is also simpler to do these calculations in 
the multiplicative variables, so that we deal with algebraic polynomials. 

\subsubsection{Remarks} The $c+$ vertex comes from $w_7$. In the
multiplicative notation, it is easy to see that it is a constant
times a factor of $U$, as 
$\alpha = 1$, $\alpha^{\prime} = 1 + \frac{1}{2}$, 
$\beta  = n$, $ \beta^{\prime} = n - \frac{1}{2}$. 

Also notice that there are two types of vertices: 
In vertices $\{w_1, $$w_2, $$w_3, $$w_4\}$, particles flow into 
the vertex, then flow out with no change in colour. Their weights 
have degree 2 in each rapidity variable. 
In vertices $\{w_5, $$w_6, $$w_7, $$w_8\}$, particles scatter with 
a change of colour. Their weights have degrees in each rapidity 
variable that vary depending on the flowing colours, and are at 
most 2.

We consider the top horizontal row, and show that the degree 
of the partition function as a polynomial in $X^{-1}$ is $2L-2$ 
(where degree is the difference of highest and lowest powers 
in $X^{-1}$).

%
\begin{center}
\begin{minipage}{4.9in}
\setlength{\unitlength}{0.0008cm}
\begin{picture}(4800, 4000)(-0100,2500)
\thicklines

\path(2400,5400)(2400,4200)
 \put(2100,5400){$1$}
 \put(2100,4200){$ $}

\path(3000,5400)(3000,4200)
 \put(2700,5400){$1$}
 \put(2700,4200){$ $}

\path(3600,5400)(3600,4200)
 \put(3300,5400){$1$}
 \put(3300,4200){$ $}

\path(4200,5400)(4200,4200)
 \put(3900,5400){$1$}
 \put(3900,4200){$ $}

\path(4800,5400)(4800,4200)
 \put(4500,5400){$1$}
 \put(4500,4200){$ $}


\path(1800,4800)(5400,4800)
 \put(1800,4900){$1$}
 \put(5400,4900){$n$}

\path(2400,2934)(2400,3294)
\path(3000,2934)(3000,3294)
\path(3600,2934)(3600,3294)
\path(4200,2934)(4200,3294)
\path(4800,2934)(4800,3294)

\path(0600,4854)(1200,4854)

\whiten\path(2490,3294)(2400,3654)(2310,3294)(2490,3294)
\whiten\path(3090,3294)(3000,3654)(2910,3294)(3090,3294)
\whiten\path(3690,3294)(3600,3654)(3510,3294)(3690,3294)
\whiten\path(4290,3294)(4200,3654)(4110,3294)(4290,3294)
\whiten\path(4890,3294)(4800,3654)(4710,3294)(4890,3294)

\whiten\path(840,4764)(1200,4854)(840,4944)(840,4764)

\put(-100,4854){$x_L$}

\put(2300,2600){$y_1$}
\put(2900,2600){$y_2$}
\put(4700,2600){$y_L$}
\end{picture}

\begin{ca}
\label{top-row}
The top row of vertices.
\end{ca}

\end{minipage}
\end{center}

\bigskip

We scan the top row from left to right. The left most vertex can 
be one of the following three types:


\begin{center}
\begin{minipage}{4.9in}
\setlength{\unitlength}{0.0007cm}
\begin{picture}(4000, 5000)(-1240,-1000)


\thicklines


\path(0620,2220)(3420,2220)
 \put(-1240,2120){$x_L$}
\whiten\path(-160,2130)(0200,2220)(-160,2310)(-160,2130)
\path(-520,2220)(-160,2220)
 \put(0620,2420){$1$}
 \put(3220,2420){$1$}

 \put(0620,3620){$L_1$}


\path(2020,0820)(2020,3620)
\whiten\path(2110,40)(2020,0400)(1930,0040)(2110,0040)
\path(2020,-320)(2020,0040)
\put(1920,-720){$y_1$}
\put(2220,1020){$1$}
\put(2220,3420){$1$}


\thicklines


\path(6620,2220)(9420,2220)
\put(4760,2120){$x_L$}
\whiten\path(5840,2130)(6200,2220)(5840,2310)(5840,2130)
\path(5480,2220)(5840,2220)
\put(6620,2420){$1$}
\put(9220,2420){$\iota$}


\path(8020,0820)(8020,3620)
\whiten\path(8110,40)(8020,400)(7930,40)(8110,40)
\path(8020,-320)(8020,40)
\put(7920,-720){$y_1$}
\put(8220,1020){$\iota$}
\put(8220,3420){$1$}

 \put(6620,3620){$L_2$}


\thicklines


\path(12620,2220)(15420,2220)
\put(10760,2120){$x_L$}
\whiten\path(11840,2130)(12200,2220)(11840,2310)(11840,2130)
\path(11480,2220)(11840,2220)
\put(12620,2420){$1$}
\put(15220,2420){$n$}


\path(14020,0820)(14020,3620)
\whiten\path(14110,40)(14020,400)(13930,40)(14110,40)
\path(14020,-320)(14020,40)
\put(13920,-720){$y_1$}
\put(14220,1020){$n$}
\put(14220,3420){$1$}

 \put(12620,3620){$L_3$}
\end{picture}
\begin{ca}
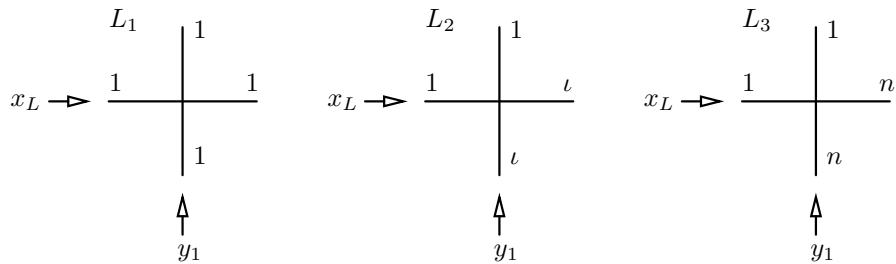
Three possible vertices on upper left corner.
\label{upperleftcorner}
\end{ca}
\end{minipage}
\end{center}

\bigskip

If it is of type $L_1$, then we are back in the original situation, 
but now with a row of length $L-1$. If it is of type $L_2$, where 
the colour $\iota$ $\in$ $\{2, 3, \cdots, n-1\}$, then we are left 
with a row of length $L-1$ of the form

%
\begin{center}
\begin{minipage}{4.9in}
\setlength{\unitlength}{0.0008cm}
\begin{picture}(4800, 4000)(-100,2500)
\thicklines

\path(3000,5400)(3000,4200)
 \put(2700,5400){$1$}
 \put(2700,4200){$ $}

\path(3600,5400)(3600,4200)
 \put(3300,5400){$1$}
 \put(3300,4200){$ $}

\path(4200,5400)(4200,4200)
 \put(3900,5400){$1$}
 \put(3900,4200){$ $}

\path(4800,5400)(4800,4200)
 \put(4500,5400){$1$}
 \put(4500,4200){$ $}


\path(2400,4800)(5400,4800)
 \put(2400,4900){$\iota$}
 \put(5400,4900){$n$}

\path(3000,2934)(3000,3294)
\path(3600,2934)(3600,3294)
\path(4200,2934)(4200,3294)
\path(4800,2934)(4800,3294)

\path(0600,4854)(1200,4854)

\whiten\path(3090,3294)(3000,3654)(2910,3294)(3090,3294)
\whiten\path(3690,3294)(3600,3654)(3510,3294)(3690,3294)
\whiten\path(4290,3294)(4200,3654)(4110,3294)(4290,3294)
\whiten\path(4890,3294)(4800,3654)(4710,3294)(4890,3294)

\whiten\path(840,4764)(1200,4854)(840,4944)(840,4764)

\put(-100,4854){$x_L$}

\put(2900,2600){$y_2$}
\put(4700,2600){$y_L$}
\end{picture}

\begin{ca}
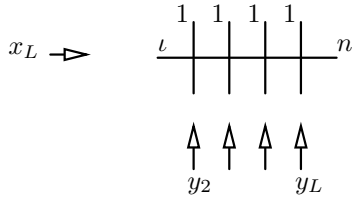
A reduced top row with colour $\iota$ flowing in from left.
\end{ca}

\end{minipage}
\end{center}

\bigskip

Since colour $\iota$ flows in from the left, but only colour $1$
flows out from the top, and $n$ flows out from right, colour 
$\iota^\prime$ must flow in at some point from below. This is 
necessary for conservation of colour. Thus we conclude that the
flow of colour $\iota$ horizontally must terminate with a vertex 
of type


\begin{center}
\begin{minipage}{4.9in}
\setlength{\unitlength}{0.0008cm}
\begin{picture}(4000, 6000)(-1280,-1000)


\thicklines


\path(0620,2220)(3420,2220)
\put(-1280,2120){$x_L$}
\whiten\path(-160,2130)(0200,2220)(-160,2310)(-160,2130)
\path(-520,2220)(-160,2220)
\put(0620,2420){$\iota$}
\put(3220,2420){$n$}


\path(2020,0820)(2020,3620)
\whiten\path(2110,0040)(2020,0400)(1930,0040)(2110,0040)
\path(2020,-320)(2020,0040)
 \put(1920,-720){$y_j$}
 \put(2220,1020){$\iota^\prime$}
 \put(2220,3420){$1$}

\end{picture}
\begin{ca}
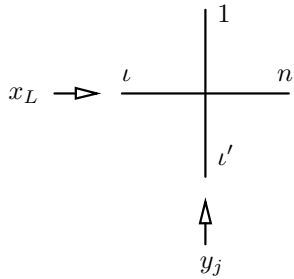
Top row vertex at which horizontally flowing colour 
$\iota$ $\in$ $\{2, 3, \cdots, n-1\}$ terminates.
$\iota + \iota^{\prime} = n+1$.
\label{iota-terminates}
\end{ca}
\end{minipage}
\end{center}

\bigskip

This introduces colour $n$ horizontally, and the latter must flow
all the way to the right end.

The third possibility is that the upper left vertex is of type $L_3$, 
colour $n$ flows at that point into the row, and propagates all the 
way to the right end where it exits.

From the above (and noting that the symmetrizing factors cancel
between pairs of vertices of types $v_5$ and $v_6$ that are allowed 
to appear in the top row), one concludes that the partition function 
as a polynomial in $X^{-1}$ is $(2L-2)$, where degree is the difference 
between highest and lowest powers of $X^{-1}$, so that an overall factor 
of $X$ from the $c+$ vertex does not contribute to the degree counting.

\subsubsection{RHS} It is straightforward to compute the degree 
of the determinant expression as a polynomial in any (multiplicative) 
variable, such as $X^{-1}$. $N$ has degree $6L$. $D$ has degree $4L-4$. 
Naive counting indicates that the determinant has degree $-4$. If 
that's correct, the degree of the determinant expression would be 
$6L - 4L + 4 - 4 = 2L$, which would disagree with that of the LHS. 
Happily, due to the $2\times 2$ block structure of the determinant, 
there is a cancellation of the leading, degree $-6$, term, thus the 
determinant has degree $-5$, and we end up once again with a degree 
$(2L-1)$ polynomial.

\subsection{Left corner recursion relations}

\subsubsection{LHS}

Consider the set of all possible vertices that can occur on the 
upper left corner. They are shown in figure \ref{upperleftcorner}.

Set $-x_L + y_1 + n-2= 0$. This forces the upper left vertex to 
be a $c+$ vertex, as the weights of the two other possibilities 
is 0. 

This also freezes the rest of the top row vertices to be of type 
2, and the rest of the first column vertices to be of type 2.

The partition function reduces to an $(L-1)\times (L-1)$ partition 
function times a constant factor of

\begin{equation}
\ll \pl_{1 \leq i \leq L-1} w_{3}(-x_i + y_1) \rr
\ll w_{c+}                                    \rr
\ll \pl_{2 \leq j \leq L  } w_{3}(-x_L + y_j) \rr 
\label{leftconstant}
\end{equation}

%
\begin{center}
\begin{minipage}{4.9in}
\setlength{\unitlength}{0.0008cm}
\begin{picture}(4800, 6000)(0, 0)
\thicklines


\path(2400,1800)(2400,5400)
\path(2420,1800)(2420,4800)
\path(2440,1800)(2440,4800)
\path(2460,1800)(2460,4800)
\path(2480,1800)(2480,4800)
 \put(2100,5400){$1$}
 \put(2100,1800){$n$}

\path(3000,4500)(3000,5400)
 \put(2700,5400){$1$}

\path(3600,4500)(3600,5400)
 \put(3300,5400){$1$}


\path(4200,4500)(4200,5400)
 \put(3900,5400){$1$}


\path(4800,4500)(4800,5400)
 \put(4500,5400){$1$}


\path(1800,4800)(5400,4800)
\path(2400,4780)(5400,4780)
\path(2400,4760)(5400,4760)
\path(2400,4740)(5400,4740)
\path(2400,4720)(5400,4720)
 \put(1800,4900){$1$}
 \put(5400,4900){$n$}

\path(1800,4200)(2700,4200)
 \put(1800,4300){$1$}

\path(1800,3600)(2700,3600)
 \put(1800,3700){$1$}

\path(1800,3000)(2700,3000)
 \put(1800,3100){$1$}

\path(1800,2400)(2700,2400)
 \put(1800,2500){$1$}
%
\path(0600,4254)(1200,4254)
\path(2400,0654)(2400,1254)
\path(3000,0654)(3000,1254)
\path(3600,0654)(3600,1254)
\path(4200,0654)(4200,1254)
\path(4800,0654)(4800,1254)
\path(0600,2454)(1200,2454)
\path(0600,3054)(1200,3054)
\path(0600,3654)(1200,3654)
\path(0600,4854)(1200,4854)
\whiten\path(2490,894)(2400,1254)(2310,894)(2490,894)
\whiten\path(3090,894)(3000,1254)(2910,894)(3090,894)
\whiten\path(3690,894)(3600,1254)(3510,894)(3690,894)
\whiten\path(4290,894)(4200,1254)(4110,894)(4290,894)
\whiten\path(4890,894)(4800,1254)(4710,894)(4890,894)
\whiten\path(840,2364)(1200,2454)(840,2544)(840,2364)
\whiten\path(840,2964)(1200,3054)(840,3144)(840,2964)
\whiten\path(840,3564)(1200,3654)(840,3744)(840,3564)
\whiten\path(840,4164)(1200,4254)(840,4344)(840,4164)
\whiten\path(840,4764)(1200,4854)(840,4944)(840,4764)
\put(0100,4854){$x_L$}
\put(0100,3054){$x_2$}
\put(0100,2454){$x_1$}
\put(2300,0250){$y_1$}
\put(2900,0250){$y_2$}
\put(4700,0250){$y_L$}
\end{picture}

\begin{ca}
\label{upper-left-corner}
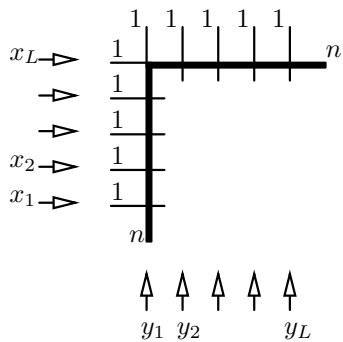
Frozen upper left corner, upper row and left column. 
Colour $1$ flows in the thin lines.
Colour $n$ flows in the thick lines.
\end{ca}

\end{minipage}
\end{center}

\bigskip

\subsubsection{RHS}

Setting $-x_L + y_1 + n-2= 0$ in the determinant expression,
and using the condition \ref{condition}, we obtain a determinant 
expression for the $(L-1)\times (L-1)$ DW partition function times 
the factor in equation \ref{leftconstant}. 

\subsection{Right corner recursion relations}

\subsubsection{LHS}

The set of all allowed vertices on the upper right corner are

%

\begin{center}
\begin{minipage}{4.9in}
\setlength{\unitlength}{0.0007cm}
\begin{picture}(4000, 6000)(-1280,-1000)


\thicklines


\path(0620,2220)(3420,2220)
\put(-1280,2120){$x_L$}
\whiten\path(-160,2130)(0200,2220)(-160,2310)(-160,2130)
\path(-520,2220)(-160,2220)
\put(0620,2420){$n$}
\put(3220,2420){$n$}


\path(2020,0820)(2020,3620)
\whiten\path(2110,40)(2020,400)(1930,40)(2110,40)
\path(2020,-320)(2020,40)
\put(1920,-720){$y_L$}
\put(2220,1020){$1$}
\put(2220,3420){$1$}

 \put(3420,3620){$R_1$}


\thicklines


\path(6620,2220)(9420,2220)
\put(4720,2120){$x_L$}
\whiten\path(5840,2130)(6200,2220)(5840,2310)(5840,2130)
\path(5480,2220)(5840,2220)
 \put(6620,2420){$\iota$}
 \put(9220,2420){$n$}


\path(8020,0820)(8020,3620)
\whiten\path(8110,40)(8020,400)(7930,40)(8110,40)
\path(8020,-320)(8020,40)
\put(7920,-720){$y_L$}
\put(8220,1020){$\iota^\prime$}
\put(8220,3420){$1$}

 \put(9420,3620){$R_2$}

\thicklines


\path(12620,2220)(15420,2220)
\put(10720,2120){$x_L$}
\whiten\path(11840,2130)(12200,2220)(11840,2310)(11840,2130)
\path(11480,2220)(11840,2220)
\put(12620,2420){$1$}
\put(15220,2420){$n$}


\path(14020,0820)(14020,3620)
\whiten\path(14110,0040)(14020,0400)(13930,0040)(14110,0040)
\path(14020,-320)(14020,0040)
 \put(13920,-720){$y_L$}
 \put(14220,1020){$n$}
 \put(14220,3420){$1$}

 \put(15420,3620){$R_3$}

\end{picture}
\begin{ca}
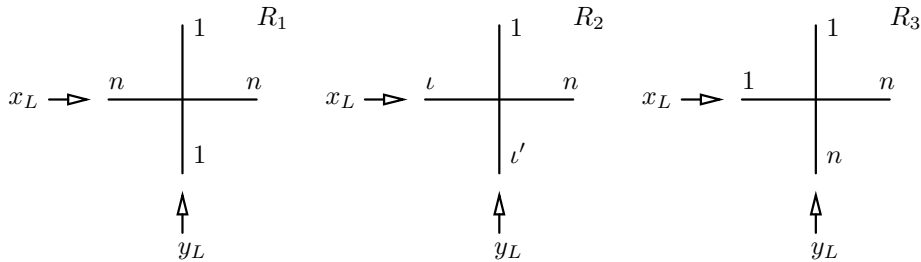
Three possible vertices on upper right corner.
\label{upperrightcorner}
\end{ca}
\end{minipage}
\end{center}

\bigskip

%
\begin{center}
\begin{minipage}{4.9in}
\setlength{\unitlength}{0.0008cm}
\begin{picture}(4800, 6000)(-0100, 0)
\thicklines


\path(2400,5400)(2400,4500)
 \put(2100,5400){$1$}

\path(3000,5400)(3000,4500)
 \put(2700,5400){$1$}

\path(3600,5400)(3600,4500)
 \put(3300,5400){$1$}

\path(4200,5400)(4200,4500)
 \put(3900,5400){$1$}

\path(4800,5400)(4800,1800)
\path(4780,4800)(4780,1800)
\path(4760,4800)(4760,1800)
\path(4740,4800)(4740,1800)
\path(4720,4800)(4720,1800)
 \put(4500,5400){$1$}
 \put(4900,1800){$n$}


\path(1800,4800)(5400,4800)
\path(4800,4800)(5400,4800)
\path(4800,4780)(5400,4780)
\path(4800,4760)(5400,4760)
\path(4800,4740)(5400,4740)
\path(4800,4720)(5400,4720)

 \put(1800,4900){$1$}
 \put(5400,4900){$n$}

\path(4500,4200)(5400,4200)
\path(4500,4180)(5400,4180)
\path(4500,4160)(5400,4160)
\path(4500,4140)(5400,4140)
\path(4500,4120)(5400,4120)
 \put(5400,4300){$n$}

\path(4500,3600)(5400,3600)
\path(4500,3580)(5400,3580)
\path(4500,3560)(5400,3560)
\path(4500,3540)(5400,3540)
\path(4500,3520)(5400,3520)
 \put(5400,3700){$n$}

\path(4500,3000)(5400,3000)
\path(4500,2980)(5400,2980)
\path(4500,2960)(5400,2960)
\path(4500,2940)(5400,2940)
\path(4500,2920)(5400,2920)
 \put(5400,3100){$n$}

\path(4500,2400)(5400,2400)
\path(4500,2380)(5400,2380)
\path(4500,2360)(5400,2360)
\path(4500,2340)(5400,2340)
\path(4500,2320)(5400,2320)
 \put(5400,2500){$n$}
%
\path(0600,4254)(1200,4254)
\path(2400,0654)(2400,1254)
\path(3000,0654)(3000,1254)
\path(3600,0654)(3600,1254)
\path(4200,0654)(4200,1254)
\path(4800,0654)(4800,1254)
\path(0600,2454)(1200,2454)
\path(0600,3054)(1200,3054)
\path(0600,3654)(1200,3654)
\path(0600,4854)(1200,4854)
\whiten\path(2490,894)(2400,1254)(2310,894)(2490,894)
\whiten\path(3090,894)(3000,1254)(2910,894)(3090,894)
\whiten\path(3690,894)(3600,1254)(3510,894)(3690,894)
\whiten\path(4290,894)(4200,1254)(4110,894)(4290,894)
\whiten\path(4890,894)(4800,1254)(4710,894)(4890,894)
\whiten\path(840,2364)(1200,2454)(840,2544)(840,2364)
\whiten\path(840,2964)(1200,3054)(840,3144)(840,2964)
\whiten\path(840,3564)(1200,3654)(840,3744)(840,3564)
\whiten\path(840,4164)(1200,4254)(840,4344)(840,4164)
\whiten\path(840,4764)(1200,4854)(840,4944)(840,4764)
\put(-0100,4854){$x_L$}
\put(-0100,3054){$x_2$}
\put(-0100,2454){$x_1$}
\put(2300,0250){$y_1$}
\put(2900,0250){$y_2$}
\put(4700,0250){$y_L$}
\end{picture}

\begin{ca}
\label{upper-right-corner}
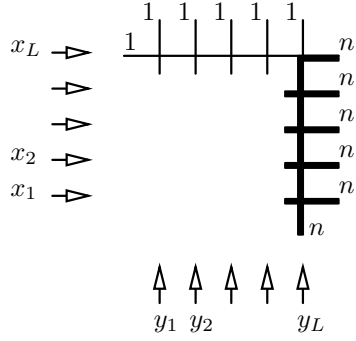
Frozen upper right corner, upper row and right column. Colour $1$
flows in the thin lines. Colour $n$ flows in the thick lines.
\end{ca}

\end{minipage}
\end{center}

\bigskip

Setting $-x_L + y_L = 0$, we dictate that the upper right vertex
is of type $c+$, as the weights of all other possible vertices
vanish.

This also reduces the rest of the upper row vertices to be of
type 1, and the rest of the final column vertices to be of type
1.

This reduces the $L\times L$ DW partition function to an
$(L-1)\times (L-1)$ DW partition function, times a constant 
factor

\begin{equation}
\ll \pl_{1 \leq j \leq L-1} w_1(-x_L + y_j) \rr
\ll w_{c+}                                  \rr
\ll \pl_{1 \leq i \leq L-1} w_1(-x_i + y_L) \rr 
\label{rightconstant}
\end{equation}

\subsubsection{RHS}

Setting $-x_L + y_L = 0$ in the determinant expression, and using 
the condition \ref{condition}, we obtain a determinant expression 
for the $(L-1)\times (L-1)$ DW partition function times the factor 
in equation \ref{rightconstant}. However, this reduction requires 
the use of condition \ref{condition}.

\subsection{Initial condition}

\subsubsection{LHS} By construction, the $1\times 1$ DW partition 
function is the $c+$ vertex.

\subsubsection{RHS} Using condition \ref{condition}, one can show 
by direct calculation, and using the condition \ref{condition}, 
that, for $L=1$, the determinant expression reduces to the weight 
of the $c+$ vertex.

\section{The homogeneous Limit}\label{hom-lim}

Taking the homogeneous limit of the $so(n)$ DW partition function, 
following the footsteps of \cite{izergin}, is straightforward. For 
convenience, we re-write $M^{2\times 2}_{2L\times 2L, ij}$ as

\begin{equation}
M^{2\times 2}_{2L\times 2L, ij}=
\ll
\begin{array}{ll}
\Phi(-x_{i}+y_{j})&\Phi(-x_{i}+y_{j}+1)\\ 
\Phi(-x_{i}+y_{j}-1)&\Phi(-x_{i}+y_{j})\\ 
\end{array}
\rr 
\nonumber
\end{equation}

\noindent where $\Phi(x)=<x><x+1>$. Following \cite{izergin} and 
\cite{cfk}, we set $x_{1}=x$, consider the limits $x_{i}\rightarrow x$, 
expand in $x_{i}$ around $x$ to order $i-1$, and subtract lower double 
rows successively from higher ones to obtain

\begin{multline}
\frac{
{\ll 
\pl_{j=1}^{L}
\pl_{p=0}^{2} 
[- x + y_j + p][- x + y_j + p - 1]
\rr}^{L}
det \ll M^{2\times 2, \  {\rm semi-hom}}_{2L\times 2L}\rr
}
{
\pl_{i=1}^{L-1}{\ll i! \rr}^2
\pl_{1\leq i<j \leq L}
\pl_{p=1}^{2} [-y_i + y_j + p][- y_i + y_j + p - 1]} \\
\label{1hom}
\end{multline}

\noindent where $\Phi^{(n)}(x)$ is the $n$-th derivative of $\Phi$ 
with respect to its argument, and 

\begin{equation}
M^{2\times 2, \ {\rm semi-hom}}_{2L\times 2L, ij} = 
\ll
\begin{array}{ll}
\Phi^{(i-1)}(-x+y_{j})&\Phi^{(i-1)}(-x+y_{j}+1)\\
\Phi^{(i-1)}(-x+y_{j}-1)&\Phi^{(i-1)}(-x+y_{j})\\ 
\end{array}
\rr 
\nonumber
\end{equation}

Applying similar arguments to the vertical rapidities, and combining 
results, we obtain the following expression for the homogeneous 
$L\times L$ $so(n)$ DW partition function

\begin{equation}
Z^{2\times 2 {\rm hom}}_{L\times L}= 
\pl_{i=1}^{L-1}
{\ll i!  \rr}^{-4}
{\ll
\pl_{p=0}^{2} [-x+y+p][-x+y+p-1]
\rr}^{L^2}
det \ll M^{2\times 2, \  {\rm hom}}_{2L\times 2L}\rr
\label{2hom}
\end{equation}

\noindent where  
$M^{2\times 2, \  {\rm hom}}_{L\times L, ij}$ is 

\begin{align}
\ll
\begin{array}{llll}
\Phi^{(i+j-2)}(-x+y)&\Phi^{(i+j-2)}(-x+y+1)\\ 
\Phi^{(i+j-2)}(-x+y-1)&\Phi^{(i+j-2)}(-x+y)\\ 
\end{array}
\rr 
\nonumber
\end{align}

\section{Remarks}

The original motivation for considering DW boundary conditions, 
in the context of the six vertex model
\cite{korepin,izergin,korepin-book}, was that the corresponding
partition function is a basic building block in the algebraic
Bethe ansatz approach to correlation functions. Subsequently, 
and unexpectedly, it turned out that these partition functions 
are related to enumerating alternating sign matrices
\cite{bressoud-book}. It is still unclear whether the DW
partition functions obtained in \cite{cfk} and in this work are 
the objects required for analogous computations in higher spin 
and in $so(n)$ vertex models, respectively, or in generalised 
enumeration problems. We hope to address these computations in 
subsequent publications. 

Condition \ref{condition} has been indispensable. It is unclear 
whether our result is a special case of a more general result 
that applies to all values of the crossing parameter $\lambda$.

For vertex models based on affine $su(n)$, there are no analogues 
of minimal and maximal colours, as the $Z_{n+1}$ symmetry of the
affine $su(n)$ Dynkin diagram puts all colours on equal footing. 
Choosing any two colours $\{\kappa_1, \kappa_2\}$ as boundary 
colours to obtain the analogue of DW boundary conditions, one 
finds that the affine $su(n)$ quantum $R$ matrix restricts the
allowed vertices in such a way that only these two colours 
$\{\kappa_1, \kappa_2\}$ are allowed to flow on the inner
segments of the partition function. This effectively restricts 
the allowed configurations to those of an affine $su(2)$
sub-algebra of the affine $su(n)$ under consideration, and one
ends up with the same determinant expression as Izergin's.

For affine $C_n$ vertex models, and models based on twisted 
affine algebras, the $2\times 2$ block structure that we used 
in our proposal \ref{proposal} does not satisfy the required 
initial condition, and one needs a more elaborate proposal, 
but that is beyond the scope of this work.

Finally, we wish to remark that the fact that DW boundary
conditions allow for $2L$ conditions that are able to determine 
polynomials of degree $2L-1$ indicates that the proof used in
this work applies to models with partition functions that are 
one degree higher in each rapidity variable.

\section*{Acknowledgements} We wish to thank A~Caradoc, J~de~Gier, 
N~Kitanine and T~A~Welsh for discussions. This work was supported 
by a summer scholarship from the Department of Mathematics and 
Statistics, U of Melbourne, and by the Australian Research Council 
(ARC).


\end{document}